\documentclass[sigconf]{acmart}
\AtBeginDocument{%
  }

\usepackage{booktabs}
\usepackage{multirow}
\usepackage{graphicx}
\usepackage{subfigure}
\usepackage{algorithm}
\usepackage{algorithmicx}
\usepackage{algpseudocode}
\usepackage{amsmath}
\usepackage[most]{tcolorbox}
\usepackage{xcolor}

\algnewcommand\algorithmicparam{\textbf{Parameter:}}
\algnewcommand\Param{\item[\algorithmicparam]} 

\newtcolorbox{ragprompt}{
  colback=white,
  colframe=black,
  boxrule=0.8pt,
  arc=6pt,
  outer arc=6pt,
  left=6pt, right=6pt, top=6pt, bottom=6pt,
  fontupper=\small\ttfamily,
  breakable,
  enhanced,
  before skip=10pt,
  after skip=10pt
}

\newcommand{\promptsep}[1]{%
  \begin{center}
    \vspace{-0.5em}
    ======================= #1 =======================
    \vspace{-0.5em}
  \end{center}%
}

\setcopyright{acmlicensed} \copyrightyear{2026} \acmYear{2026}
\acmDOI{XXXXXXX.XXXXXXX}

\acmISBN{978-1-4503-XXXX-X/2018/06}

\begin{document}

\title{CQC-RAG: Robust Retrieval-Augmented Generation via Cross-Query Consistency}

\author{Yanjia Sun}
\affiliation{%
  \institution{University of Electronic Science and Technology of China}
  \city{Chengdu}
  \state{Sichuan}
  \country{China}
}
\email{sunyanjia@std.uestc.edu.cn}

\author{Sifan Liu}
\affiliation{%
  \institution{University of Electronic Science and Technology of China}
  \city{Chengdu}
  \state{Sichuan}
  \country{China}
} \email{liusifan@std.uestc.edu.cn}

\author{Jie Shao}
\authornote{Corresponding author: Jie Shao.}
\affiliation{%
  \institution{University of Electronic Science and Technology of China}
  \city{Chengdu}
  \state{Sichuan}
  \country{China}
}
\email{shaojie@uestc.edu.cn}

\renewcommand{\shortauthors}{Sun et al.}

\begin{abstract}
Retrieval-Augmented Generation (RAG) has become a common approach
for improving the factuality of Large Language Models (LLMs), yet
its reliability remains highly sensitive to how external evidence is
retrieved and used. Semantically equivalent queries with different
syntactic forms may lead to different retrieval results, while
irrelevant or misleading documents can further induce hallucinated
answers. Existing multi-path reasoning methods improve robustness by
sampling multiple candidate answers and applying voting- or
confidence-based selection, but they still face two limitations:
diversity is often injected through uncontrollable decoding
randomness, and answer evaluation is usually confined to a single
query-induced evidence view. To address these limitations, we
propose a Cross-Query Consistency Hypothesis: correct answers tend
to maintain high confidence across semantically equivalent but
syntactically diverse queries, whereas noise-induced hallucinations
exhibit unstable confidence under such query variations. Based on
this hypothesis, we introduce CQC-RAG, a framework that co-designs
query-level diversity injection with cross-query consistency
evaluation. CQC-RAG rewrites the original question into diverse but
meaning-preserving queries, reranks a shared document pool to
construct query-conditioned reasoning contexts, applies an
evidence-grounded protocol to extract answer-evidence pairs and
selects answers according to their confidence stability across these
contexts. This design enables self-evaluation without external
supervision and does not rely on expanded retrieval coverage.
Experiments on four open-domain question answering benchmarks show
that CQC-RAG outperforms the strongest previous multi-query baseline
by +4.76 pp EM on TriviaQA and +9.12 pp EM on MuSiQue, validating
the effectiveness of cross-query consistency for filtering
noise-induced hallucinations. Code is available at
\url{https://github.com/FrancesPlus/CQC-RAG}.
\end{abstract}




\maketitle

\section{Introduction}
\label{sec:intro}

Retrieval-Augmented Generation (RAG) has become a widely adopted
paradigm for mitigating hallucinations and knowledge obsolescence in
Large Language Models (LLMs) by retrieving external knowledge and
incorporating it into the generation context
\cite{DBLP:conf/nips/LewisPPPKGKLYR020,DBLP:journals/jmlr/IzacardLLHPSDJRG23}.
Despite its effectiveness, RAG remains fragile in practice. On the
one hand, the specific formulation of a query significantly affects
retrieval quality: semantically equivalent queries that differ only
in syntactic structure, such as active versus passive voice or
synonym substitution, lead to an average of 15\%--20\% degradation
in retrieval performance across different datasets, with extreme
cases exceeding 40\% \cite{DBLP:conf/eacl/CaoBYAS26}. On the other
hand, retrieved documents may contain information irrelevant to the
question, and studies have shown that such irrelevant inputs
substantially interfere with the reasoning process of models,
causing errors and hallucinations
\cite{DBLP:journals/corr/abs-2601-07226}. These observations suggest
that answer reliability in RAG depends not only on the language
model, but also on how the question induces and prioritizes external
evidence.

A common strategy for improving robustness is to generate multiple
reasoning paths and then select a final answer from them.
Self-Consistency \cite{DBLP:conf/iclr/0002WSLCNCZ23} samples
multiple reasoning paths from the same query and selects the most
consistent answer through majority voting, establishing the
theoretical foundation that correct answers tend to converge across
diverse reasoning paths while incorrect answers are scattered
stochastically. SPARC-RAG \cite{DBLP:journals/corr/abs-2602-00083}
uses adaptive sequential-parallel reasoning, while Speculative RAG
\cite{DBLP:conf/iclr/00020LZMP0MTSLP25} adopts a draft-then-verify
mechanism to produce and evaluate candidate answers. These methods
improve robustness by exploiting answer agreement across multiple
generations. Yet, when applied to open-domain RAG settings without
external supervision, existing methods still face two fundamental
limitations:
\begin{itemize}
\item \textbf{Challenge 1: Uncontrollability of the diversity
injection mechanism.} Existing approaches rely primarily on
token-level temperature sampling, which often struggles with a
trade-off between semantic drift (at high temperatures) and path
redundancy (at low temperatures). More importantly, because all
generated paths share the same initial query formulation, this
decoding-level randomness fails to address the inherent sensitivity
of retrieval systems to query formulations
\cite{DBLP:conf/eacl/CaoBYAS26}.
\item \textbf{Challenge 2: Limited discriminability of answer evaluation.}
Conventional majority voting treats all reasoning paths equally, and
even recent confidence-weighted selections remain restricted to
repeat inference over a single retrieval perspective, leaving them
vulnerable to errors consistently reinforced by biased or noisy
evidence
\cite{DBLP:conf/acl/TaubenfeldSOFGG25,DBLP:journals/corr/abs-2502-18581}.
\end{itemize}

These two challenges are tightly coupled. The quality of diversity
injection directly determines the distributional characteristics of
candidate answers, which in turn affects the difficulty and accuracy
of evaluation. Conversely, the effectiveness of the evaluation
mechanism determines whether the value of diversity injection can be
fully exploited. Therefore, diversity injection and answer
evaluation must be co-designed as an integrated whole.

To this end, we propose CQC-RAG, a Cross-Query Consistency RAG
framework that co-designs diversity injection and answer evaluation
as an integrated system. CQC-RAG is built on the \textbf{Cross-Query
Consistency Hypothesis}, which posits that correct answers exhibit
stable high confidence across semantically equivalent but
syntactically diverse queries due to their grounding in genuinely
relevant documents. In contrast, noise-induced hallucinated answers
rely on spurious, context-dependent correlations, causing
significant confidence fluctuations across these query variations.
To operationalize this hypothesis, CQC-RAG first generates a set of
semantically equivalent but syntactically diverse queries through
query rewriting, introducing controlled diversity at the query level
rather than at the decoding level. All rewritten queries share the
same retrieved document pool, but a reranker reorders the documents
according to the semantic relevance to each individual query,
producing differentiated priority orderings across queries that
constitute distinct reasoning contexts. The model performs parallel
reasoning over these contexts, generating diverse candidate answers.
Finally, we compute the confidence of each answer under each query
perspective based on model logits and select the answer that
maintain high confidence across queries, characterized by high mean
and low variance, achieving answer self-evaluation without external
supervision signals. It is important to emphasize that all queries
share the same document pool, and the performance gains come from
the cross-query consistency framework rather than from expanded
retrieval coverage.

Through this pipeline, diversity injection and answer evaluation are
seamlessly coupled, as query-level variations do not act as random
decoding noise but rather systematically construct the structured,
multi-perspective contexts required for our consistency-based
evaluation.

We validate the effectiveness of CQC-RAG on open-domain Question
Answering (QA) datasets including TriviaQA and MuSiQue. Experimental
results demonstrate that CQC-RAG significantly outperforms baseline
methods such as Self-Consistency and Speculative RAG, indicating
that cross-query consistency evaluation can effectively identify and
filter hallucinated answers induced by noise, thereby validating the
Cross-Query Consistency Hypothesis.

Our technical contributions are summarized as follows:
\begin{itemize}
\item We propose the Cross-Query Consistency Hypothesis, which establishes
a theoretical connection between the stability of answer confidence
across diverse query formulations and answer correctness.
\item We design CQC-RAG, a framework that co-designs query-level diversity
injection with cross-query consistency evaluation. By reranking a
shared document pool with semantically equivalent but syntactically
diverse queries to produce differentiated reasoning contexts,
CQC-RAG enables answer verification across multiple knowledge
perspectives without increasing the retrieval budget.
\item Extensive experiments on single-hop and multi-hop
question-answering benchmarks demonstrate the superior effectiveness
and efficiency of our method.
\end{itemize}

\begin{figure*}[t]
  \centering
  \includegraphics[width=0.86\linewidth]{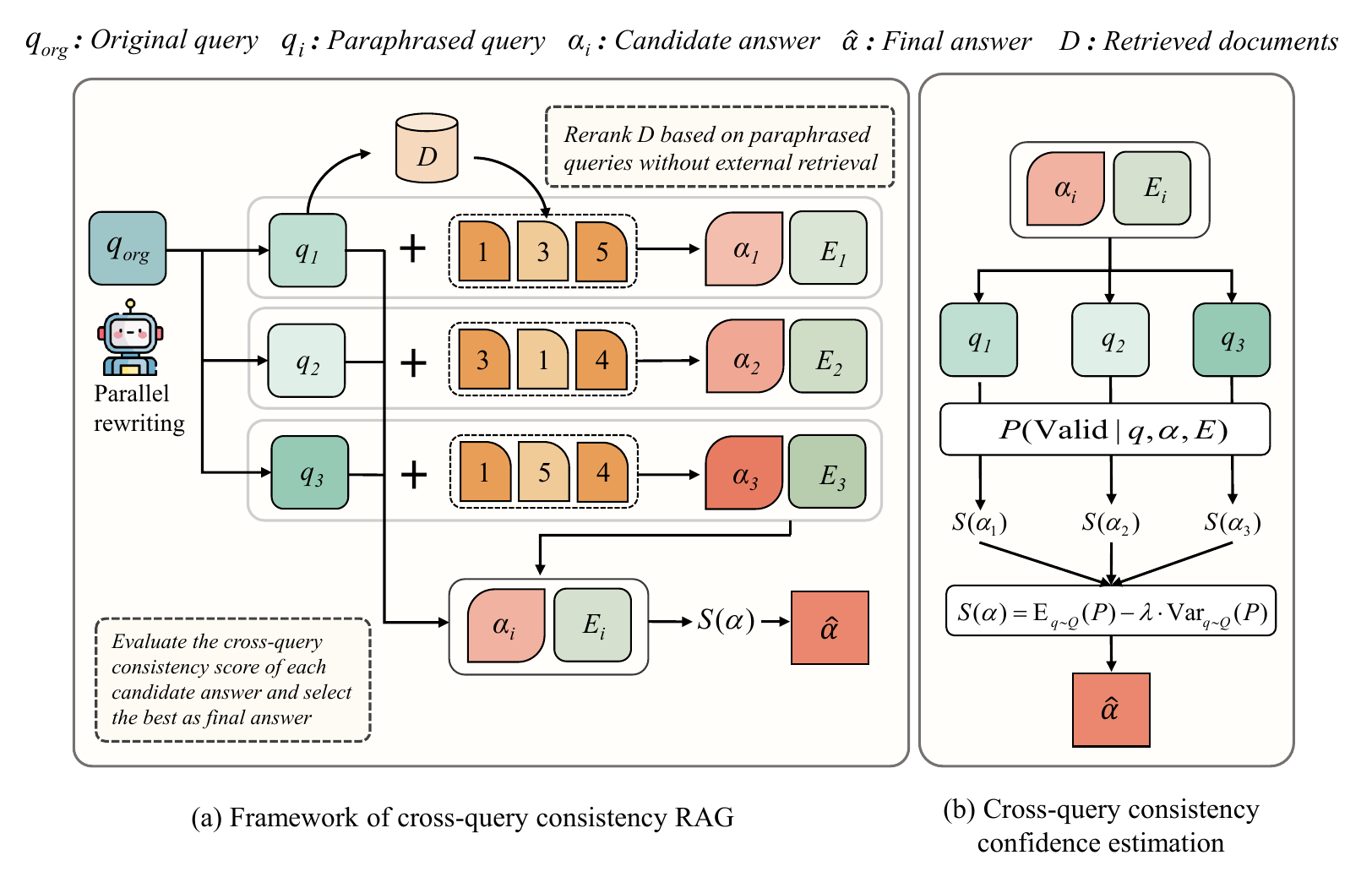}
  \caption{Illustration of CQC-RAG. CQC-RAG employs a parallel rewriting
mechanism to construct multiple queries that are semantically
equivalent yet syntactically diverse. This enables multi-perspective
reasoning over the context, yielding several candidate answers. The
optimal answer is subsequently selected based on cross-query
consistency confidence estimation.}
  \label{fig:main}
\end{figure*}

\section{Related Work}

Our proposed CQC-RAG framework lies at the intersection of improving
generation robustness and evaluating answer reliability.
Specifically, our work is most closely related to two lines of
research: (1) injecting query-level diversity to enhance the
robustness of RAG systems against contextual noise; (2) leveraging
multi-path reasoning and self-consistency to mitigate model
hallucinations and designing unsupervised evaluation mechanisms for
response verification. We review these closely related works in
detail below.

\subsection{Query-Level Diversity for Robust RAG}

Retrieval-Augmented Generation (RAG) enhances LLMs by incorporating
external documents into inference
\cite{DBLP:conf/nips/LewisPPPKGKLYR020,DBLP:journals/jmlr/IzacardLLHPSDJRG23},
but this also makes generation sensitive to how evidence is
retrieved and organized. Irrelevant passages can induce
over-reasoning and hallucinations
\cite{DBLP:journals/corr/abs-2404-03302}, while semantically
equivalent query reformulations can substantially change retrieval
recall and downstream answers \cite{DBLP:conf/eacl/CaoBYAS26}. This
suggests that robust RAG requires not only better evidence, but also
a more controlled way to expose evidence under query variation.

Existing query optimization methods address this issue mainly from a
retrieval perspective. Query rewriting methods such as
Rewrite-Retrieve-Read \cite{DBLP:journals/corr/abs-2305-14283} and
Adaptive-RAG \cite{DBLP:conf/naacl/JeongBCHP24} employ LLMs to
reformulate queries into more canonical forms, aiming to predict a
single optimal entry point for retrieval. SiReRAG
\cite{DBLP:conf/iclr/ZhangCFBZMXW25} constructs enhanced indexing
mechanisms that capture both semantically similar and explicitly
related information for multi-hop reasoning. RARE
\cite{DBLP:conf/acl/TranY0WZHO025} and DMQR-RAG
\cite{DBLP:journals/corr/abs-2411-13154} use iterative or
multi-query rewriting to improve coverage and precision. These
methods show the utility of query reformulation, but they generally
use diversity to enlarge, refine, or merge retrieved evidence.

Therefore, these improved methods do not directly resolve the first
challenge: how to inject controllable diversity for reasoning rather
than merely for retrieval. Once rewritten queries are merged or
optimized toward a single best query, their query-specific evidence
priorities are largely discarded; meanwhile, token-level sampling
still inherits the same query-induced retrieval bias. Our work
instead treats query diversity as a controlled way to construct
complementary reasoning contexts over a shared document pool,
leaving the detailed design to the method section.

\subsection{Discriminative Answer Evaluation}

The second challenge concerns how to select a reliable answer once
multiple candidate reasoning paths are available. Self-Consistency
\cite{DBLP:conf/iclr/0002WSLCNCZ23} samples multiple reasoning paths
from the same query and chooses the most frequent answer,
establishing the principle that correct answers tend to converge
while incorrect ones are more scattered. This paradigm has inspired
RAG-oriented extensions. SPARC-RAG
\cite{DBLP:journals/corr/abs-2602-00083} adaptively allocates
sequential and parallel reasoning with specialized agents, while
Speculative RAG \cite{DBLP:conf/iclr/00020LZMP0MTSLP25} follows a
draft-then-verify pipeline in which candidate answers are generated
from document subsets and then scored by a verifier. However,
producing more candidates does not by itself guarantee that the
final selector can distinguish genuinely supported answers from
plausible errors.

Recent work improves selection mechanisms by using richer signals
than raw majority. Taubenfeld et al.
\cite{DBLP:conf/acl/TaubenfeldSOFGG25} show that confidence-weighted
voting can reduce the number of required samples by exploiting the
model's own confidence signal, and emphasize the importance of
within-question discrimination. Ranked Voting
\cite{DBLP:conf/acl/WangWH25} replaces simple frequency counting
with rank-based scores to better preserve high-quality minority
answers. Self-Certainty \cite{DBLP:journals/corr/abs-2502-18581}
measures the divergence between the predicted token distribution and
a uniform distribution, showing that correct and incorrect answers
can form separable clusters in confidence space. DistriVoting
\cite{DBLP:journals/corr/abs-2603-03872} further models trajectory
confidence as a mixture distribution and links the separation
between confidence modes to voting accuracy. These findings
collectively suggest that the model's internal confidence signals
contain rich information for distinguishing correct from incorrect
answers within the same query context.

Nevertheless, they still largely evaluate answers under a single
query-conditioned evidence view. When biased or noisy evidence is
prioritized under a query-conditioned context, wrong answers may
appear plausible or confident, leaving the second challenge
unresolved: the selector lacks a discriminative axis beyond repeated
inference over the same retrieval perspective. CQC-RAG addresses
this limitation through cross-query consistency evaluation, which
tests answer reliability beyond a single query-conditioned evidence
view.

\section{Methodology}

In this section, we present the CQC-RAG framework.
Figure~\ref{fig:main} and Algorithm~\ref{alg:cqc_rag} provide an
overview of the proposed CQC-RAG.

\begin{algorithm}[t] \small
    \caption{Framework of CQC-RAG}
    \label{alg:cqc_rag}
    \begin{algorithmic}[1]
    \Require Original query $q_{org}$, Retrieved documents $\mathcal{D}$
    \Param Paraphrase count $N$, Penalty coefficient $\lambda(\cdot)$, Models $\{M_{\text{rewrite}}, M_{\text{rerank}}, M_{\text{reason}}, M_{\text{eval}}\}$
    \Ensure Final answer $\hat{\alpha}$
    \Statex \textbf{Stage 1: Parallel Rewriting}
    \State $\mathcal{Q}_{para} \leftarrow M_{\text{rewrite}}(q_{org}, N)$ \Comment{Generate diverse queries under constraints}
    \State $\mathcal{Q} \leftarrow \{q_{org}\} \cup \mathcal{Q}_{para}$ \Comment{Form the multi-view query set}
    \Statex \textbf{Stage 2: Multi-view Reranking \& Reasoning}
    \State Initialize candidate set $\mathcal{A} \leftarrow \emptyset$
    \For{each $q_i \in \mathcal{Q}$ \textbf{in parallel}}
        \State $C_i \leftarrow M_{\text{rerank}}(q_i, \mathcal{D})$ \Comment{Obtain view-specific context}
        \State $\mathcal{A} \leftarrow \mathcal{A} \cup \{ M_{\text{reason}}(C_i, q_i) \}$ \Comment{Generate pair $(\alpha_i, E_i)$ via strict protocol}
    \EndFor
    \Statex \textbf{Stage 3: Candidate Selection}
    \For{each candidate $(\alpha_k, E_k) \in \mathcal{A}$}
        \State $P_{scores} \leftarrow \emptyset$
        \For{each query view $q_j \in \mathcal{Q}$}
            \State $(\ell_{\text{Yes}}, \ell_{\text{No}}) \leftarrow M_{\text{eval}}(\text{Valid} | \alpha_k, E_k, q_j)$
            \State $P_{scores}.\text{append}(\frac{\exp(\ell_{\text{Yes}})}{\exp(\ell_{\text{Yes}}) + \exp(\ell_{\text{No}})})$ \Comment{Normalize logits}
        \EndFor
        \State $\mu_k \leftarrow \text{Mean}(P_{scores})$
        \State $S(\alpha_k) \leftarrow \mu_k - \lambda(\mu_k) \cdot \text{Var}(P_{scores})$
    \EndFor
    \State \textbf{return} $\hat{\alpha} \leftarrow \arg\max_{\alpha_k} S(\alpha_k)$
    \end{algorithmic}
\end{algorithm}

We begin by introducing some preliminaries, followed by the query
rewriting mechanism designed to generate semantically equivalent but
syntactically diverse query variants. We then describe the
cross-query reasoning and evidence extraction process. Finally, we
detail the cross-query consistency confidence estimation strategy
for candidate answer selection.

\subsection{Preliminaries}

For tasks that heavily rely on external knowledge, a standard RAG
instance can be formalized as the triplet $(q, \mathcal{D},
\alpha)$. Here, $q$ denotes the input query; $\mathcal{D}$
represents the retrieved external knowledge corpus containing large
amounts of factual information; and $\alpha$ denotes the expected
correct answer. The primary goal of an RAG system is to retrieve a
relevant contextual subset $C= \{c_1, c_2, ..., c_m\} \subset
\mathcal{D}$ associated with $q$, and leverage the informational
content within to formulate a coherent and accurate response
$\alpha$.

In our proposed framework, we denote the user's original query as
$q_{org}$. Through the parallel rewriting mechanism, we generate a
set of semantically isomorphic paraphrased queries, denoted as
$\mathcal{Q}_{para} = \{q_1, q_2,..., q_N\}$. Since the original
query $q_{org}$ is also involved during inference, the entire set of
queries utilized for reasoning is denoted as $\mathcal{Q} =
\{q_{org}, q_1, q_2,..., q_N\}$. For each query in this set, the
model performs parallel inference to yield a corresponding set of
candidate answers and evidence spans, denoted as $\mathcal{A} =
\{\alpha_i\}_{i=1}^{|\mathcal{Q}|}$. Our objective is to identify
the final answer $\hat{\alpha}$ from $\mathcal{A}$ that exhibits the
highest cross-query consistency.

\subsection{Query Rewriting for Controlled Diversity}

Existing multi-path reasoning methods, exemplified by
Self-Consistency \cite{DBLP:conf/iclr/0002WSLCNCZ23}, derive
diversity from temperature sampling during decoding. All reasoning
paths share the same query and the same reasoning context. As
discussed in Section~\ref{sec:intro}, this approach suffers from a
fundamental limitation: the diversity is uncontrollable (high
temperature risks unfaithfulness while low temperature yields
redundant paths), and it cannot address the sensitivity of retrieval
outcomes to query formulation \cite{DBLP:conf/eacl/CaoBYAS26}.
Multi-query rewriting approaches such as DMQR-RAG
\cite{DBLP:journals/corr/abs-2411-13154} generate diverse query
variants across multiple information levels (information-equivalent,
information-expanded, and information-reduced) to improve retrieval
coverage. However, DMQR-RAG merges all retrieved documents into a
single pool for one-pass generation, using query diversity solely to
expand the document pool rather than to produce differentiated
reasoning contexts for independent inference.

Our approach takes a distinct perspective: we leverage query
rewriting not to expand retrieval coverage, but to produce multiple
differentiated reasoning contexts based on the same fixed document
pool $\mathcal{D}$ via query-specific reranking, enabling
cross-query answer verification. The key insight, formalized as the
Cross-Query Consistency Hypothesis in Section~\ref{sec:intro}, is
that a correct answer should maintain high confidence across diverse
query perspectives, while a hallucinated answer supported by
spurious or noisy evidence should exhibit instability under query
variation. This verification objective imposes a specific
requirement on the rewriting strategy: all rewritten queries must be
strictly semantically equivalent to $q_{org}$, targeting the same
factual answer. Unlike DMQR-RAG's information-level rewrites that
deliberately alter the semantic scope to retrieve broader or more
focused documents, our rewriting preserves strict semantic
equivalence because any semantic drift between queries would
invalidate the premise of cross-query consistency. Answers generated
from queries with different semantic scopes are not meaningfully
comparable for consistency evaluation.

We formalize this reasoning process as follows. Let $I$ denote the
user's latent semantic intent and let $P(q \mid I)$ represent the
distribution over possible query formulations that express the same
intent $I$. Since the reasoning context is determined by the query
through reranking, we write $C(q)$ for the top-$k$ context produced
by reranking $\mathcal{D}$ with respect to $q$. The probability of
answer $\alpha$ given intent $I$ can then be expressed as:
\begin{equation}
  P(\alpha \mid I) = \int P(\alpha \mid C(q),\, q) \; P(q \mid I) \; dq.
\end{equation}
Since this integral is intractable in the discrete text space, we
approximate it by averaging over the finite sample set $\mathcal{Q}$
of rewritten queries, each treated as a sample from $P(q \mid I)$:
\begin{equation}
P(\alpha \mid I) \approx \frac{1}{|\mathcal{Q}|} \sum_{q_i \in
\mathcal{Q}} P(\alpha \mid C_i,\, q_i),
\end{equation}
where $C_i = C(q_i)$ denotes the query-specific reasoning context
obtained by reranking $\mathcal{D}$ with respect to $q_i$. Unlike
standard ensemble methods where all members share the same input,
each $q_i$ here produces its own reasoning context $C_i$ from the
shared document pool. This distinction is operationally critical:
the averaging process attenuates the spurious, context-dependent
correlations that may accidentally support a hallucinated answer
under a single query formulation, while preserving semantic signals
that are consistently supported across multiple reasoning contexts.

To maximize the quality of this approximation, the paraphrased query
set $\mathcal{Q}_{para}$ must satisfy two requirements: (1) strict
semantic equivalence with the user's intent $I$, ensuring all
queries target the same factual answer; and (2) maximal syntactic
diversity, ensuring different queries induce different document
orderings and thus different reasoning contexts. We implement this
through a few-shot prompting strategy governed by two dimensions of
constraints:
\begin{enumerate}
\item \textbf{Hard constraints.} Given that any alteration of named
entities may lead to retrieval drift, we instruct the LLM to
strictly freeze all named entities, preserving their literal forms
to ensure all generated $q_i$ are anchored to the same factual
baseline.
\item \textbf{Soft constraints.} We restructure $q_{org}$ across three
dimensions to maximize syntactic diversity while maintaining
semantic equivalence:
\begin{itemize}
\item \textit{Lexical perturbation}: Synonym substitution for non-entity
predicates to resolve lexical mismatch between queries and
documents, thereby increasing the likelihood that different document
passages are promoted during reranking.
\item \textit{Syntactic restructuring}: Alteration of the syntactic tree
structure through active/passive voice switching and clause
reordering, producing queries that emphasize different aspects of
the same intent.
\item \textit{Pragmatic modal shift}: Mixing interrogative forms with
imperative instructions. This hybrid strategy produces queries that
stylistically align with diverse document genres, accommodating
passages written in different rhetorical modes.
\end{itemize}
\end{enumerate}

\subsection{Cross-Query Reasoning}

Given the rewritten query set $\mathcal{Q} = \{q_{org}, q_1, q_2,
\ldots, q_N\}$, we perform cross-query reranking and reasoning to
translate the query-level diversity established in the previous
stage into differentiated reasoning outcomes.

\textbf{Query-specific reranking.} For each query $q_i \in
\mathcal{Q}$, we apply a reranking model $M_{\text{rerank}}$ over
the shared initial document pool $\mathcal{D}$ to construct a
query-specific reasoning context $C_i = M_{\text{rerank}}(q_i,
\mathcal{D})$. Since the reranker scores each document based on its
semantic relevance to the specific query $q_i$, syntactically
different queries naturally produce different document orderings and
thus different top-$k$ contexts. Crucially, no new documents are
introduced during this process and the only variation is in which
passages are promoted to the top-$k$ positions. This design isolates
the effect of document prioritization from retrieval coverage: any
performance gains of CQC-RAG over single-query baselines cannot be
attributed to seeing more documents, but rather to the cross-query
consistency mechanism that operates on the resulting variation in
reasoning contexts.

\textbf{Evidence-grounded reasoning.} We distribute the
query-specific reasoning contexts $\{C_i\}$ and their corresponding
queries to the reasoning model $M_{\text{reason}}$ in parallel. To
enforce faithfulness to the source text and mitigate hallucinations,
we implement a strict evidence-grounded reasoning protocol: the LLM
is explicitly instructed to first locate and extract the specific
segment from $C_i$ that answers the query, defined as the supporting
evidence span $E_i$. Subsequently, the model is constrained to
derive the final answer $\alpha_i$ based solely on $E_i$. All
$|\mathcal{Q}|$ reasoning paths execute in parallel, enabling the
framework to scale with minimal latency overhead. This process
yields a set of reasoning tuples $\{(\alpha_i,
E_i)\}_{i=1}^{|\mathcal{Q}|}$. These tuples serve as the candidate
pool for cross-query consistency evaluation.

\subsection{Cross-Query Consistency Confidence Estimation}
\label{sec:CQC_confidence}

The core contribution of CQC-RAG lies in shifting the paradigm of
answer evaluation from a single reasoning context to cross-query
consistency. Existing confidence estimation approaches---whether
based on majority voting frequency
\cite{DBLP:conf/iclr/0002WSLCNCZ23}, confidence-weighted voting
\cite{DBLP:conf/acl/TaubenfeldSOFGG25}, token distribution certainty
\cite{DBLP:journals/corr/abs-2502-18581}, or distributional
filtering \cite{DBLP:journals/corr/abs-2603-03872}---evaluate
candidates within the same query's reasoning context and cannot
reliably detect hallucinations reinforced by biased or noisy
evidence under one particular query formulation.

For any candidate answer $\alpha$ and its supporting evidence span
$E$, we define the cross-query consistency score $S(\alpha)$ in
Eq.~\eqref{eq:score}:
\begin{equation}
S(\alpha) = \underbrace{\mathbb{E}_{q \sim \mathcal{Q}}
[P(\text{Valid} \mid \alpha, E, q)]}_{\text{Semantic Consensus}} -
\lambda(\mu) \cdot \underbrace{\text{Var}_{q \sim \mathcal{Q}}
[P(\text{Valid} \mid \alpha, E, q)]}_{\text{Cross-Query
Instability}}, \label{eq:score}
\end{equation}
where $\mu = \mathbb{E}_{q \sim \mathcal{Q}} [P(\text{Valid} \mid
\alpha, E, q)]$ denotes the mean validation score. The evaluation is
performed over answer-evidence pairs $(\alpha, E)$ rather than over
the full reasoning context $C_i$: since $E$ distills the relevant
information from $C_i$ into a focused evidence span, the evaluator
can assess whether the evidence semantically supports the answer
under each query perspective without being confounded by irrelevant
passages. Importantly, since $E$ is a direct extraction of factual
content from the source documents, its validity as supporting
evidence does not depend on the specific syntactic form of the
query---any semantically equivalent query formulation should
recognize the same factual statement as valid support.

To compute $P(\text{Valid} \mid \alpha, E, q)$, we leverage the
auto-regressive nature of the evaluator model $M_{\text{eval}}$.
Rather than generating a free-form judgment (which would introduce
additional variance from the generation process itself), we perform
a single forward pass over a verification prompt and extract the
logits corresponding to the ``Yes'' and ``No'' tokens
($\ell_{\text{Yes}}, \ell_{\text{No}}$), normalizing them over a
restricted binary space:
\begin{equation}
P(\text{Valid} \mid \alpha, E, q) =
\frac{\exp(\ell_{\text{Yes}})}{\exp(\ell_{\text{Yes}}) +
\exp(\ell_{\text{No}})}.
\end{equation}
This logits-based evaluation offers two advantages over
generation-based alternatives: (1) it eliminates the stochasticity
of open-ended generation, producing deterministic confidence
estimates that enable meaningful cross-query comparison; and (2) the
restricted binary normalization mitigates the acquiescence bias
where models disproportionately favor affirmative responses during
open-ended generation.

To further penalize candidates that achieve artificially inflated
scores through superficial matching on a minority of views, we
design the penalty coefficient to be adaptive: $\lambda(\mu) =
\lambda_0 \cdot \mu$, where $\lambda_0$ is a base hyperparameter.
This dynamic mechanism imposes stricter variance constraints (higher
penalty) on candidates with high mean scores, ensuring their overall
rank is driven by genuine cross-query robustness rather than being
skewed by a few extreme outlier scores. Conversely, it provides
necessary tolerance for candidates with generally lower confidence.

To mitigate the tendency of models to rate their own generations
higher than others \cite{DBLP:conf/nips/ZhengC00WZL0LXZ23}, in our
experiments we employ a separate LLM distinct from the inference
model to perform this confidence evaluation task.

\section{Experiments}

\subsection{Datasets and Evaluation Metrics}

To demonstrate the effectiveness of CQC-RAG, we evaluate CQC-RAG on
four open-domain question answering benchmarks spanning diverse
retrieval challenges:

(1) \textbf{TriviaQA} \cite{DBLP:conf/acl/JoshiCWZ17} is a
large-scale dataset that requires factual knowledge retrieval. We
evaluate our method on a standard open-domain split containing 2,000
queries. The associated retrieval corpus is constructed from a
comprehensive Wikipedia dump, and candidate passages are retrieved
using a dense retriever.

(2) \textbf{PopQA} \cite{DBLP:conf/acl/MallenAZDKH23} is a long-tail
knowledge dataset. To explicitly evaluate the model's reliance on
external knowledge, we use its challenging long-tail subset, which
comprises 1,399 queries. In this subset, the target entities receive
fewer than 100 monthly Wikipedia page views, making parametric
memory insufficient and external retrieval essential.

For both single-hop benchmarks, we adopt the same experimental
settings as \cite{DBLP:conf/iclr/AsaiWWSH24}.

(3) \textbf{HotpotQA} \cite{DBLP:conf/nips/GutierrezS0Y024} and (4)
\textbf{MuSiQue} \cite{DBLP:journals/tacl/TrivediBKS22} are two
multi-hop QA datasets involving comparison, synthesis, and
bridge-type reasoning across multiple documents. Following the
standard multi-hop evaluation setting established in previous work
\cite{DBLP:conf/nips/GutierrezS0Y024}, we randomly sample 1,000
queries from the validation set of each dataset and extract all
candidate passages (both supporting facts and distractors) from the
original annotations of the sampled queries. This
distractor-inclusive corpus setting rigorously tests the model's
ability to identify multi-hop reasoning paths amid highly confusing
but topically related information.

We adopt two complementary metrics following standard open-domain QA
evaluation practices:
\begin{itemize}
\item \textbf{Exact Match (EM)}: The predicted answer string is normalized
and compared against all acceptable ground-truth answers after the
same normalization. A prediction scores 1 if the normalized
prediction exactly equals any normalized ground-truth, and 0
otherwise. The maximum score across all ground-truth variants is
taken for each sample.
\item \textbf{F1-score (F1)}: The normalized prediction and ground-truth
are tokenized by whitespace. Precision is computed as the fraction
of prediction tokens appearing in the ground-truth, recall as the
fraction of ground-truth tokens appearing in the prediction, and
F1-score is the harmonic mean. Again, the maximum over all
ground-truth variants is reported.
\end{itemize}
We report the average of EM and F1 as a combined indicator.

\subsection{Baselines}

We compare CQC-RAG with some widely-used baseline methods.
\begin{enumerate}
\item \textbf{Standard RAG.} We evaluate three representative
instruction-tuned LLMs under a standard retrieve-then-generate
pipeline: Mistral-7B-Instruct
\cite{DBLP:journals/corr/abs-2310-06825}, Qwen3-8B
\cite{DBLP:journals/corr/abs-2505-09388}, and LLaMA-3.1-8B-Instruct
\cite{DBLP:journals/corr/abs-2407-21783}. Each model retrieves the
top-$k$ documents and generates answers with greedy decoding.
\item \textbf{Multi-path RAG.} These methods generate multiple candidate
answers and apply diverse selection strategies. Self-Certainty
\cite{DBLP:journals/corr/abs-2502-18581} computes a
KL-divergence-based certainty score over the full output vocabulary
distribution and selects the answer with the highest certainty.
DMQR-RAG \cite{DBLP:journals/corr/abs-2411-13154} generates multiple
query variants across different information levels, merges the
retrieved documents from all variants into a single pool, and
performs one-pass generation on the merged context. Speculative RAG
\cite{DBLP:conf/iclr/00020LZMP0MTSLP25} clusters the retrieved
documents into diverse subsets and generates multiple draft answers
via a lightweight drafting model, then employs a larger verifier
model to select the best draft based on self-reflection scores.
SPARC-RAG \cite{DBLP:journals/corr/abs-2602-00083} implements an
adaptive sequential-parallel multi-agent framework where specialized
agents handle query rewriting, context management, answer
generation, and answer evaluation across iterative rounds. We
evaluate its unfinetuned version using Qwen3-8B as the backbone for
all agent components.
\end{enumerate}
For a fair comparison, all multi-path baselines are evaluated with
the same backbone model and retrieval infrastructure as CQC-RAG,
while their candidate counts and sampling temperatures follow the
configurations reported in the original papers.

\subsection{Experimental Settings}

\subsubsection{Implementation Details}

We implement CQC-RAG using Qwen3-8B as $M_{\text{reason}}$ (with
thinking mode disabled for fair comparison) and
Mistral-7B-Instruct-v0.2 as $M_{\text{eval}}$. For query rewriting,
we set temperature=0.7 to ensure sufficient syntactic diversity; for
reasoning, we use greedy decoding (temperature=0) to produce
deterministic answers. The logits-based evaluation is inherently
deterministic. We set $N=4$ (determined by the sensitivity analysis)
and $\lambda_0=0.5$ as default hyperparameters. For the reranker
model, we use BGE-reranker-v2-m3
\cite{DBLP:journals/corr/abs-2402-03216} to obtain the top-$k$
query-specific context for each query variant. We set the reranking
cutoff $k=5$ for single-hop datasets (TriviaQA, PopQA) and $k=10$
for multi-hop datasets (MuSiQue, HotpotQA), since multi-hop
questions require synthesizing evidence across multiple documents
and benefit from a larger reasoning context. This top-$k$ setting is
applied uniformly to both CQC-RAG and all baselines to ensure a fair
comparison.

\subsubsection{Reproducibility}

The only source of randomness in CQC-RAG is the query rewriting
stage (temperature=0.7). For the main comparison
(Table~\ref{tab:performance_comparison_all}), we report results
averaged over 5 random seeds (42, 123, 456, 789, 1024) with standard
deviation to ensure statistical robustness. More details regarding
parallel rewriting can be found in Appendix. For analysis
experiments, we report results from a single representative run to
isolate the effect of individual components; the relative
performance trends are consistent across seeds. All other pipeline
components are fully deterministic.

\subsection{Main Results}

\begin{table*}[t]
    \centering
    \caption{Performance comparison on the single-hop (TriviaQA, PopQA) and multi-hop (MuSiQue, HotpotQA) datasets using EM, F1, and average scores.}
    \label{tab:performance_comparison_all}
    \resizebox{\textwidth}{!}{
    \setlength{\tabcolsep}{6pt}
    \begin{tabular}{lcccccccccccc}
        \toprule
        \multirow{3}{*}{\bf Method}
        & \multicolumn{6}{c}{\bf Single-hop}
        & \multicolumn{6}{c}{\bf Multi-hop} \\
        \cmidrule(lr){2-7} \cmidrule(lr){8-13}
        & \multicolumn{3}{c}{\bf TriviaQA} & \multicolumn{3}{c}{\bf PopQA}
        & \multicolumn{3}{c}{\bf MuSiQue} & \multicolumn{3}{c}{\bf HotpotQA} \\
        \cmidrule(lr){2-4} \cmidrule(lr){5-7} \cmidrule(lr){8-10} \cmidrule(lr){11-13}
        & EM & F1 & AVG. & EM & F1 & AVG. & EM & F1 & AVG. & EM & F1 & AVG. \\
        \midrule

        \multicolumn{13}{c}{\it Standard RAG} \\
        \midrule
        Mistral-7B-Instruct & 39.13 & 54.21 & 46.67 & 25.73 & 36.79 & 31.26 & 3.80 & 14.13 & 8.97 & 6.60 & 16.62 & 11.61 \\
        Llama-3.1-8B-Instruct & 54.35 & 64.40 & 59.38 & 42.96 & 50.35 & 46.66 & 14.50 & 21.22 & 17.86 & 16.50 & 23.19 & 19.85 \\
        Qwen3-8B & 49.18 & 63.61 & 56.40 & 40.67 & 49.59 & 45.13 & 12.30 & 24.32 & 18.31 & 12.60 & 25.63 & 19.12 \\
        \midrule
        \multicolumn{13}{c}{\it Multi-path RAG} \\
        \midrule
        Self-Certainty & 51.10 $\pm$ 0.40 & 61.30 $\pm$ 0.40 & 56.20 & 42.00 $\pm$ 0.20 & 48.00 $\pm$ 0.30 & 45.00 & 9.40 $\pm$ 0.10 & 15.20 $\pm$ 0.30 & 12.30 & 52.20 $\pm$ 0.40 & 65.00 $\pm$ 0.30 & 58.60 \\
        Speculative RAG & 54.44 & 64.03 & 59.24 & 37.88 & 42.29 & 40.09 & 14.90 & 20.56 & 17.73 & 33.00 & 42.05 & 37.53 \\
        DMQR-RAG & \underline{54.89} & \underline{66.04} & \underline{60.47} & \underline{46.46} & \textbf{53.31} & \underline{49.89} & 23.50 & 33.92 & 28.71 & \textbf{54.70} & \textbf{68.56} & \textbf{61.63} \\
        SPARC-RAG & 36.68 & 47.34 & 42.01 & 30.38 & 40.59 & 35.49 & \underline{30.50} & \underline{41.78} & \underline{36.14} & 42.20 & 58.99 & 50.60 \\
        \midrule
        \textbf{CQC-RAG} & \textbf{59.65 $\pm$ 0.35} & \textbf{70.73 $\pm$ 0.29} & \textbf{65.19} & \textbf{46.86 $\pm$ 0.28} & \underline{53.19 $\pm$ 0.22} & \textbf{50.03} & \textbf{32.62 $\pm$ 0.30} & \textbf{44.13 $\pm$ 0.25} & \textbf{38.38} & \underline{52.50 $\pm$ 0.59} & \underline{67.60 $\pm$ 0.44} & \underline{60.05} \\
        \bottomrule
    \end{tabular}
    }
\end{table*}

\textbf{CQC-RAG achieves state-of-the-art overall performance.}
CQC-RAG attains the highest Avg on three out of four datasets,
including both single-hop benchmarks (TriviaQA: 65.19\%; PopQA:
50.03\%) and the multi-hop (MuSiQue: 38.38\%). On TriviaQA, CQC-RAG
achieves +4.72 percentage points Avg advantage over DMQR-RAG, and on
MuSiQue the improvement over SPARC-RAG reaches +2.24 percentage
points Avg. Compared with its own backbone model Qwen3-8B under
standard RAG, CQC-RAG improves TriviaQA EM by +10.47 percentage
points and MuSiQue EM by +20.32 percentage points, demonstrating
that our cross-query consistency mechanism extracts substantially
more reasoning capability from the identical retrieval
infrastructure and base LLM.

\textbf{Superiority over multi-path selection methods.}
Self-Certainty, which applies confidence-based selection over
temperature-sampled candidates from the same reasoning context,
achieves modest gains on single-hop datasets (TriviaQA EM 51.10\%).
However, its performance severely degrades on MuSiQue (EM 9.40\%),
where complex multi-hop queries amplify retrieval noise that
single-perspective confidence signals fail to detect. CQC-RAG
circumvents this epistemic blind spot by verifying answer stability
based on cross-query consistency, delivering a massive +23.22
percentage points EM improvement over Self-Certainty on MuSiQue.

\textbf{Advantage over multi-query expansion.} DMQR-RAG also employs
query rewriting but merges all retrieved documents into a single
pool for one-pass generation, leveraging query diversity solely to
maximize retrieval coverage. In contrast, CQC-RAG outperforms
DMQR-RAG in EM across TriviaQA (+4.76 percentage points), MuSiQue
(+9.12 percentage points), and PopQA (+0.40 percentage points).
Because both methods share the exact same retrieval infrastructure
and CQC-RAG does not expand the document budget (all rewritten
queries rerank the same fixed pool), these gains are strictly
attributable to the cross-query consistency evaluation rather than
broader document recall. This is further corroborated on PopQA: both
methods achieve near-identical F1 scores, indicating comparable
semantic coverage, yet CQC-RAG's higher EM highlights its superior
precision in exact answer extraction.

\textbf{Robustness against multi-agent reasoning.} SPARC-RAG employs
iterative multi-agent reasoning with adaptive depth and width
control. While it achieves competitive performance on MuSiQue, its
efficacy drops precipitously on single-hop tasks (TriviaQA: 36.68\%
EM; PopQA: 30.38\% EM). We attribute this to the tendency of its
unfinetuned multi-agent pipeline to generate comprehensive but
excessively verbose responses, severely penalizing exact-match
evaluation on datasets demanding concise entity extraction.
CQC-RAG's evidence-grounded protocol and cross-query consistency
scoring provide more effective precision control, achieving strong
performance across both single-hop and multi-hop tasks without such
precision degradation.

\textbf{Analysis of HotpotQA performance.} HotpotQA is the only
dataset where CQC-RAG ranks second (Avg 60.05\%), trailing DMQR-RAG.
We attribute this gap to HotpotQA's structural reliance on document
titles as bridge entities: questions frequently require matching
specific titles, where the retrieval signal is title matching rather
than content matching. Since our rewriting preserves entities and
modifies only content-level syntax, it produces limited variation in
title-level relevance, reducing the effectiveness of query-specific
reranking on this dataset. Notably, DMQR-RAG's information-expanded
queries can introduce additional title-relevant terms that directly
improve title-level retrieval, explaining its advantage on this
particular dataset structure.

\subsection{Ablation Study and Selection Strategy Analysis}

\subsubsection{Component Ablation}

\begin{table*}[t]
    \centering
    \caption{Ablation study of CQC-RAG in the rewriting, reasoning and evaluation stages on TriviaQA and MuSiQue.}
    \label{tab:ablation}
    \resizebox{\textwidth}{!}{
    \begin{tabular}{lcccccc}
    \toprule
    \multirow{2}{*}{\textbf{Method}} & \multicolumn{3}{c}{\textbf{TriviaQA}} & \multicolumn{3}{c}{\textbf{MuSiQue}} \\
    \cmidrule(lr){2-4} \cmidrule(lr){5-7}
    & EM & F1 & AVG. & EM & F1 & AVG. \\
    \midrule
    \textbf{CQC-RAG} & \textbf{60.39} & \textbf{71.57} & \textbf{65.98} & \textbf{32.90} & 44.13 & \underline{38.52} \\
    \midrule
    \multicolumn{7}{l}{\textit{Rewriting Stage}} \\
    \quad w/o Rewriting & \underline{59.87} \ (-0.52 pp) & \underline{70.97} \ (-0.60 pp) & \underline{65.42} \ (-0.56 pp) & 32.50 \ (-0.40 pp) & \underline{44.43} \ (+0.30 pp) & 38.47 \ (-0.05 pp) \\
    \quad w/o Query-specific Reranking & 59.15 \ (-1.24 pp) & 69.75 \ (-1.82 pp) & 64.45 \ (-1.53 pp) & 32.50 \ (-0.40 pp) & \textbf{46.10} \ (+1.97 pp) & \textbf{39.30} \ (+0.78 pp) \\
    \multicolumn{7}{l}{\textit{Reasoning Stage}} \\
    \quad w/o Evidence-grounded Protocol & 44.47 \ (-15.92 pp) & 58.91 \ (-12.66 pp) & 51.69 \ (-14.29 pp) & 24.30 \ (-8.60 pp) & 37.93 \ (-6.20 pp) & 31.12 \ (-7.40 pp) \\
    \multicolumn{7}{l}{\textit{Evaluation Stage}} \\
    \quad w/o Cross-query Evaluation & 59.57 \ (-0.82 pp) & 70.93 \ (-0.64 pp) & 65.25 \ (-0.73 pp) & 31.40 \ (-1.50 pp) & 41.84 \ (-2.29 pp) & 36.62 \ (-1.90 pp) \\
    \quad w/o Separate Evaluator & 57.83 \ (-2.56 pp) & 70.08 \ (-1.49 pp) & 63.96 \ (-2.02 pp) & \underline{32.60} \ (-0.30 pp) & 43.95 \ (-0.18 pp) & 38.33 \ (-0.19 pp) \\
    \bottomrule
    \end{tabular}
    }
\end{table*}

To quantify the contribution of each CQC-RAG component, we
systematically remove or replace individual modules while keeping
the rest of the pipeline intact (Table~\ref{tab:ablation}). All
variants use $N=4$ paraphrases from a single representative run. We
organize the analysis by the three stages of CQC-RAG's pipeline:
rewriting, reasoning, and evaluation.

\textbf{Rewriting stage.} The \textit{w/o Rewriting} variant
replaces the query rewriting mechanism with temperature sampling:
the original query is used for a single reranking pass, and $N$=4
diverse answers are generated via stochastic decoding on the shared
reasoning context. The \textit{w/o Query-specific Reranking} variant
retains query rewriting but reranks only with $q_{org}$, sharing the
identical top-$k$ document ordering across all query variants.

On TriviaQA, removing rewriting reduces EM by 0.52 percentage points
and removing query-specific reranking reduces EM by 1.24 percentage
points, confirming that both components contribute to producing the
differentiated reasoning contexts that CQC-RAG relies on for
cross-query verification. The larger degradation from removing
reranking indicates that even with diverse queries, the consistency
mechanism cannot function effectively without differentiated
document orderings.

On MuSiQue, both variants show EM decreases while F1 increases. For
\textit{w/o Rewriting}, temperature sampling generates multiple
candidate answers from the same reasoning context. Since answers to
multi-hop questions typically require compositional reasoning over
multiple evidence pieces, temperature sampling may occasionally
produce answers that contain partially correct entities, attributes,
or descriptive spans, thereby improving token-level F1. However,
such answer-level diversity cannot replace the multi-path evidence
construction enabled by query rewriting. Without rewritten queries,
the model can only reason over a single retrieval context, making it
more prone to errors in final answer boundaries, entity
normalization, or compositional reasoning, which leads to a decrease
in exact match.

For \textit{w/o Query-specific Reranking}, the increase in F1 can be
attributed to the fact that candidate evidence not ranked under
query-specific constraints may contain more content that partially
overlaps with the gold answer, thereby improving partial-match F1.
However, because all rewritten query variants share the same top-$k$
document ordering determined by $q_{org}$, documents that are more
relevant to a particular rewritten query may not be promoted to
higher ranks. This weakens the alignment between each query variant
and its corresponding evidence, making it harder for the model to
produce a fully matched final answer. In other words, the F1
improvements observed in the ablated models on the multi-hop dataset
mainly reflect increased partial-match coverage rather than improved
final-answer precision.

\textbf{Reasoning stage.} The \textit{w/o Evidence-grounded
Protocol} variant removes the ``extract evidence first, then
answer'' constraint, instead directly generating answers from the
full reasoning context without structured evidence extraction. This
causes the most severe degradation across both datasets: TriviaQA EM
drops by 15.92 percentage points and MuSiQue EM by 8.60 percentage
points. Without evidence grounding, the model is unconstrained by
source text and produces fluent but unfaithful answers.

\textbf{Evaluation stage.} The \textit{w/o Cross-query Evaluation}
variant evaluates each candidate answer only against its originating
query, eliminating the cross-query verification. This reduces
MuSiQue EM by -1.50 percentage points and F1 by -2.29 percentage
points, the largest evaluation-stage degradation, confirming that
cross-query verification is most valuable when multiple evidence
fragments must be consistently validated. The \textit{w/o Separate
Evaluator} variant uses the reasoning model (Qwen3-8B) itself as the
evaluator instead of the independent Mistral-7B. This causes
TriviaQA EM to drop by -2.56 percentage points, the second-largest
overall degradation, validating that self-enhancement bias
\cite{DBLP:conf/nips/ZhengC00WZL0LXZ23} materially compromises
answer discrimination when the same model both generates and
evaluates answers. We further investigate this effect in the
evaluator model analysis below.

\subsubsection{Selection Strategy Comparison}

\begin{table}[t]
    \centering
    \caption{Comparison of the effects of different selection strategies.}
    \label{tab:comparison_selection}
    \resizebox{\linewidth}{!}{
    \begin{tabular}{lcccccc}
    \toprule
    \multirow{2}{*}{\textbf{Method}} & \multicolumn{3}{c}{\textbf{TriviaQA}} & \multicolumn{3}{c}{\textbf{MuSiQue}} \\
    \cmidrule(lr){2-4} \cmidrule(lr){5-7}
    & EM & F1 & AVG. & EM & F1 & AVG. \\
    \midrule
    Majority Voting & 58.24 & 68.41 & 63.33 & 27.00 & 40.32 & 33.66 \\

    \multicolumn{7}{l}{\textit{Confidence Score}} \\
    \quad Probability & 58.34 & 68.97 & 63.66 & \underline{28.40} & 38.70 & 33.55 \\
    \quad Perplexity & 58.14 & 68.76 & 63.45 & 27.80 & 38.26 & 33.03 \\
    \quad Yes-Probability & \underline{58.96} & \underline{70.34} & \underline{64.65} & 26.80 & 37.82 & 32.31 \\

    LLM-as-a-Judge & 57.93 & 69.82 & 63.88 & 27.10 & \underline{41.02} & \underline{34.06} \\
    \midrule
    \textbf{CQC-RAG} & \textbf{60.39} & \textbf{71.57} & \textbf{65.98} & \textbf{32.90} & \textbf{44.13} & \textbf{38.52} \\
    \bottomrule
    \end{tabular}
    }
\end{table}

To further validate the effectiveness of CQC-RAG's cross-query
consistency scoring, we fix the candidate pool produced by CQC-RAG's
rewriting and reasoning stages and apply five alternative selection
strategies, as shown in Table~\ref{tab:comparison_selection}. This
controls for the quality of candidates and tests only the selection
mechanism.

Specifically, \textit{Majority Voting} clusters semantically
equivalent answers using BGE-M3 embeddings and selects the cluster
with the most members. \textit{Probability} selects the answer with
the highest generation probability (product of per-token logprobs).
\textit{Perplexity} selects the answer with the lowest perplexity.
\textit{Yes-Probability} computes the probability of the model
generating ``Yes'' given the candidate answer, the retrieved
evidence, and its originating question---serving as a single-query
version of CQC-RAG's logits-based evaluation.
\textit{LLM-as-a-Judge} directly employs an LLM as an evaluator by
providing all candidate answers and the original question to
$M_{\text{eval}}$ (Mistral-7B), prompting it to select the best
answer through free-form generation.

\textbf{All alternative strategies perform within a narrow band.} On
TriviaQA, the five alternatives achieve Avg scores between 63.33\%
and 64.65\%, a spread of only 1.32 percentage points. CQC-RAG breaks
out of this range at 65.98\%, which is +1.33 percentage points over
the best alternative Yes-Probability. On MuSiQue, the gap is more
pronounced: alternatives range from 32.31\% to 34.06\% Avg, while
CQC-RAG achieves 38.52\%. This confirms that the cross-query
consistency dimension provides a qualitatively different and
stronger signal for answer discrimination than any single-context
metric and is especially effective for multi-hop reasoning where
noise-induced hallucinations are more prevalent.

\textbf{Confidence-based metrics cannot distinguish noise-induced
hallucinations.} Probability, Perplexity, and Yes-Probability each
evaluate answers using a single confidence signal without
cross-query comparison. Their poor MuSiQue performance (Avg
32--34\%) reveals a fundamental limitation: a hallucinated answer
may receive high confidence when it is reinforced by biased or noisy
evidence under one particular query-conditioned context, but this
signal is indistinguishable from a genuinely correct answer within a
single reasoning context. These results demonstrate that cross-query
consistency scoring effectively utilizes multi-view information to
filter stochastic noise, providing a more reliable and balanced
answer selection mechanism.

\subsection{Sensitivity Analysis of Paraphrase Count}

\begin{figure}[t]
\centering \subfigure[TriviaQA]{
\includegraphics[width=0.475\linewidth]{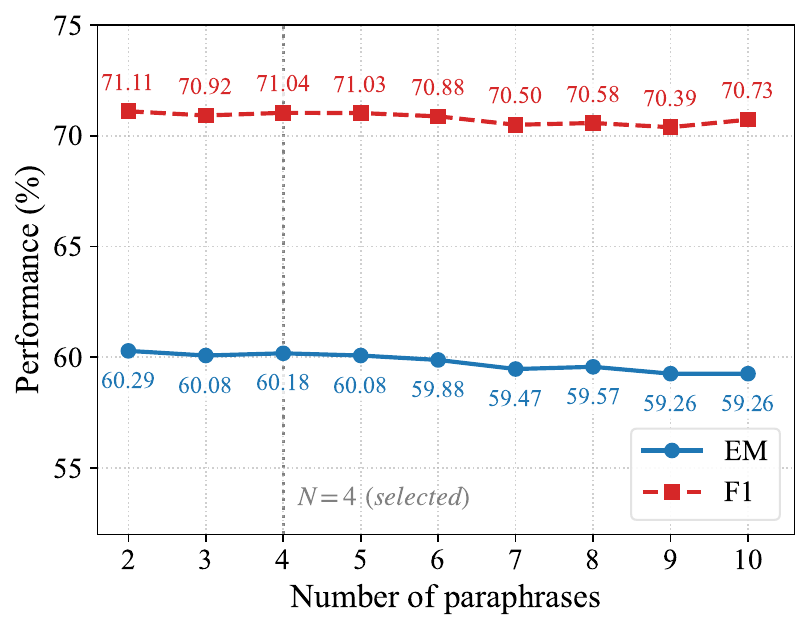}
}\\
\subfigure[MuSiQue]{
\includegraphics[width=0.475\linewidth]{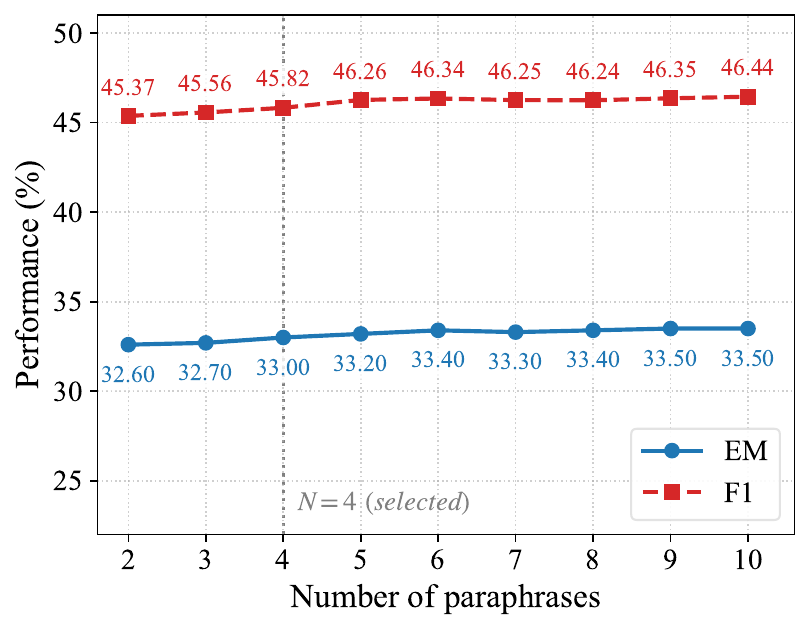}
}
\caption{Performance comparison on the TriviaQA and MuSiQue
datasets across different paraphrase counts.} \label{fig:sampling}
\end{figure}

To determine the optimal number of query rewrites and validate that
CQC-RAG's gains are not merely an artifact of increased candidate
count, we evaluate performance as a function of $N$. The results are
presented in Figure~\ref{fig:sampling}.

\textbf{Performance remains stable with a small number of
paraphrases.} On TriviaQA, the best performance is achieved at
$N=2$, with 60.29\% EM and 71.11\% F1. Increasing $N$ from 2 to 5
leads to only minor fluctuations, with EM remaining around 60\% and
F1 around 71\%. This suggests that a small set of query rewrites is
already sufficient to provide diverse reasoning contexts for
cross-query consistency verification.

\textbf{Excessive paraphrases bring limited or diminishing returns.}
On TriviaQA, performance gradually decreases when $N$ becomes
larger, dropping to 59.47\% EM at $N=7$ and 59.26\% EM at $N=10$,
indicating that additional paraphrases may introduce redundancy or
lower-quality rewrites. On MuSiQue, performance continues to improve
as $N$ increases, but the gains are relatively small: EM improves
from 33.00\% at $N=4$ to 33.50\% at $N=10$, while F1 increases from
45.82\% to 46.44\%. Thus, although more paraphrases can slightly
benefit multi-hop reasoning, the marginal improvement becomes
limited compared with the additional computational cost.

Balancing performance, robustness, and computational cost, we adopt
$N=4$ as the default configuration. Although $N=2$ yields the
highest TriviaQA score and larger values of $N$ further improve
MuSiQue slightly, $N=4$ provides a strong trade-off across both
datasets. It avoids the performance degradation observed on TriviaQA
with excessive paraphrases while capturing sufficient query
diversity for multi-hop reasoning on MuSiQue.

\subsection{Evaluator Model Analysis}

\begin{table}[t]
    \centering
    \caption{Comparison of the effects of different evaluator models.}
    \label{tab:evaluator_model}
    \resizebox{\linewidth}{!}{
    \begin{tabular}{lcccccc}
    \toprule
    \multirow{2}{*}{\textbf{Method}} & \multicolumn{3}{c}{\textbf{TriviaQA}} & \multicolumn{3}{c}{\textbf{MuSiQue}} \\
    \cmidrule(lr){2-4} \cmidrule(lr){5-7}
    & EM & F1 & AVG. & EM & F1 & AVG. \\
    \midrule
    Qwen3-8B & 58.24 & 70.23 & 64.24 & 32.60 & 43.54 & 38.07 \\
    Llama-3.1-8B-Instruct & \textbf{60.80} & \textbf{71.73} & \textbf{66.27} & \underline{32.80} & \underline{43.90} & \underline{38.35} \\
    Mistral-7B-Instruct & \underline{60.39} & \underline{71.57} & \underline{65.98} & \textbf{32.90} & \textbf{44.13} & \textbf{38.52} \\
    \bottomrule
    \end{tabular}
    }
\end{table}

As discussed in Section~\ref{sec:CQC_confidence}, we employ a
separate evaluator model $M_{\text{eval}}$ distinct from the
reasoning model $M_{\text{reason}}$ to mitigate self-enhancement
bias, a well-documented phenomenon where LLMs tend to rate their own
generated outputs more favorably than those produced by other
models. This experiment validates our evaluator design from two
complementary perspectives: the critical need for model decoupling,
and the generalizability across independent evaluator architectures.

\textbf{Necessity of a separate evaluator.} The results in
Table~\ref{tab:evaluator_model} show that replacing the distinct
$M_{\text{eval}}$ with the reasoning model itself
$M_{\text{reason}}$ degrades the TriviaQA EM to 58.24\%, which is
more than 2 percentage points lower than the decoupled
configurations. This drop confirms that self-enhancement bias
materially compromises the discriminative power of cross-query
evaluation. When the evaluator is biased towards its own reasoning
paths, it becomes overly permissive, thereby attenuating the
cross-query variance signal that CQC-RAG heavily relies upon to
distinguish genuinely supported answers from hallucinations.

\textbf{Generalizability across independent evaluators.} While
decoupling is important, CQC-RAG is highly robust to the specific
choice of the independent evaluator. To verify this, we fix
$M_{\text{reason}}$ as Qwen3-8B and employ Mistral-7B-Instruct and
Llama-3.1-8B-Instruct as two architecturally distinct,
similarly-sized evaluators. Both independent evaluators achieve
comparable and strong performance, maintaining EM fluctuations
within a narrow margin. Specifically, the difference in Exact Match
scores between the two independent evaluators is as small as 0.41
percentage points on TriviaQA and 0.10 percentage points on MuSiQue.
This robustness confirms a key property of CQC-RAG: the confidence
estimation mechanism requires only relative semantic consistency in
the evaluator's cross-query logits, rather than relying on the
absolute calibration of any specific model. When the evaluator
remains structurally independent from the reasoning model, the
framework tends to deliver robust performance.

\subsection{Rewriting Quality Analysis}

A core premise of CQC-RAG is that query rewriting produces
semantically equivalent yet syntactically diverse queries. To verify
this property and address potential concerns regarding semantic
drift, we quantitatively evaluate both the semantic consistency and
structural diversity of our rewrites. Specifically, we randomly
sample 500 queries per dataset and compute three metrics: (1)
\textbf{Semantic Similarity}, defined as the average cosine
similarity between the BGE-M3 embeddings of the original query and
its rewrites; (2) \textbf{BLEU-2 Distance}, defined as one minus the
pairwise BLEU-2 score among the rewrites to measure structural and
bigram divergence; and (3) \textbf{Jaccard Distance}, measuring the
average proportion of unshared unigrams across all pairs of
rewrites. For semantic similarity, values closer to 1.0 indicate
strict intent preservation, while for BLEU-2 Distance and Jaccard
Distance, higher values reflect greater surface-form diversity.

As shown in Table~\ref{tab:rewriting_analysis}, the average semantic
similarity consistently exceeds 0.93 across both datasets. This is
well above the 0.85 threshold established in our system design,
confirming that the applied hard constraints effectively preserve
the original user intent and prevent semantic drift. Simultaneously,
the lexical diversity metrics indicate substantial token-level
variation, with BLEU-2 Distance and Jaccard Distance averaging
between 0.50 and 0.55. This validates that our soft constraints
successfully introduce sufficient syntactic diversity, which is
crucial for producing the differentiated document orderings required
in the subsequent reranking stage.

\begin{table}[t]
    \centering
    \caption{Quantitative analysis of query rewriting quality, measuring both semantic consistency and syntactic diversity.}
    \label{tab:rewriting_analysis}
    \resizebox{\linewidth}{!}{
    \begin{tabular}{lccc}
    \toprule
    Dataset & Semantic Similarity & BLEU-2 Distance & Jaccard Distance \\
    \midrule
    TriviaQA & 0.93 $\pm$ 0.03 & 0.55 $\pm$ 0.14 & 0.53 $\pm$ 0.12  \\
    MuSiQue & 0.94 $\pm$ 0.03 & 0.50 $\pm$ 0.11 & 0.52 $\pm$ 0.09  \\
    \bottomrule
    \end{tabular}
    }
\end{table}

\begin{table*}[t] \small
    \centering
    \caption{Computational complexity and parallel latency decomposition of
multi-path RAG methods. The latency represents the theoretical
wall-clock time under sufficient parallel GPU capacity. $T_{reason}$
and $T_{rewrite}$ denote the latencies of standard reasoning and
query rewriting, respectively. $T_{draft}$ and $T_{verify}$ denote
the drafting and verification latencies of Speculative RAG. For
SPARC-RAG, $T_{\text{agents\_d}}$ represents the sequential agent
latency at iteration depth $d$. For CQC-RAG, $\epsilon$ represents
the mathematically marginal latency of query-specific reranking and
prefill-only evaluation.}
    \label{tab:overhead}
    \begin{tabular}{lcccl}
    \toprule
    \multirow{2}{*}{\textbf{Method}} & \textbf{Retrieval Stage} & \multicolumn{2}{c}{\textbf{LLM Inference Stage}} & \textbf{Parallel Latency} \\
    \cmidrule(lr){2-2} \cmidrule(lr){3-4} \cmidrule(lr){5-5}
    & Retrieval Calls & Sequential LLM Calls & Evaluation Pass & Parallel Bottleneck \\
    \midrule
    Standard RAG & 1 & 1 & None & $T_{reason}$ \\
    Self-Certainty & 1 & 1 & None & $T_{reason}$ \\
    Speculative RAG & 1 & 2 & Prefill-only Scoring & $T_{draft} + T_{verify}$ \\
    DMQR-RAG & 1--5 & 2 & None & $T_{rewrite} + T_{reason}$ \\
    SPARC-RAG & 4--16 & 12--48 & Full Autoregressive Generation & $\sum_{d=1}^{D} (T_{\text{agents\_d}})$ \\
    \midrule
    CQC-RAG & 1 & 2 & Prefill-only Scoring & $T_{rewrite} + T_{reason} + \epsilon$ \\
    \bottomrule
    \end{tabular}
\end{table*}

\subsection{Computational Overhead Analysis}

To evaluate CQC-RAG's computational cost, Table~\ref{tab:overhead}
compares representative multi-path RAG methods across three
sequential pipeline stages: retrieval routing, LLMs inference, and
theoretical parallel latency.

\textbf{Generative overhead is on par with standard multi-path
methods.} Under our default configuration of $N=4$ rewritten
queries, as determined in the sensitivity analysis, CQC-RAG requires
6 generative calls: one query-rewriting call and five independent
reasoning calls. First, in terms of the retrieval stage, CQC-RAG
requires only 1 retrieval, as all rewritten queries share a single
retrieved document pool. In contrast, other multi-query methods like
DMQR-RAG require 5 distinct retrievals, and SPARC-RAG requires 4--16
retrievals due to its iterative loops. Second, for LLM inference
stage, CQC-RAG requires only 2 stages: query rewriting and parallel
reasoning. This sequential profile is identical to DMQR-RAG and
Speculative RAG, and significantly faster than SPARC-RAG, which
requires up to 12 sequential generation steps across its multi-agent
reasoning loop. Because the 5 parallel reasoning paths in CQC-RAG
are entirely independent, modern inference engines can execute them
concurrently as a single batched inference pass, preventing
sequential latency accumulation.

\textbf{Cross-query evaluation avoids decoding bottlenecks.} The
main additional overhead of CQC-RAG comes from cross-query
evaluation. We define CQC-RAG's additional evaluation-related
overhead as $\epsilon = T_{rerank} + T_{eval\_prefill}$, where
$T_{rerank}$ represents the query-specific reranking latency over
the shared pool, and $T_{eval\_prefill}$ is the prefill-only
evaluation latency. CQC-RAG uses a compact 7B autoregressive LLM as
the evaluator, but it does not ask the model to generate full
evaluation rationales. Instead, each answer-evidence-query tuple is
processed with a single forward pass, and the binary Yes/No logits
are directly used for scoring. To run this evaluation stage
efficiently, CQC-RAG groups all 5 generated tuples into a single
batched prefill pass, allowing the evaluator to score all candidate
answers simultaneously without executing separate sequential model
calls. In comparison, while Speculative RAG also utilizes a batched
pass, it must process a larger batch of 10 drafts containing much
longer generated rationales. This prefill-only design avoids
sequential decoding during evaluation and therefore reduces the
overhead compared with generation-based judging. Because both
operations avoid autoregressive token generation, their cumulative
latency is computationally marginal.

\textbf{Overall latency and cost-reliability trade-offs.} Assuming
sufficient parallel GPU capacity, CQC-RAG's parallel latency
bottleneck is effectively $T_{rewrite} + T_{reason} + \epsilon$.
Since the evaluation latency is negligible ($\epsilon \ll
T_{reason}$), our end-to-end parallel latency is structurally close
to DMQR-RAG and avoids the sequential latency accumulation of
methods such as SPARC-RAG.

Although single-path RAG or unverified Self-Consistency may have
lower raw inference latency, they lack the cross-query verification
signal needed to filter noise-induced hallucinations. CQC-RAG
therefore introduces a limited evaluation overhead in exchange for
improved factual reliability, offering a practical cost-reliability
trade-off for scenarios where answer correctness is critical.

\section{Conclusion}

In this work, we address the reliability limitations of RAG under
single query formulation and noisy evidence. We propose the
Cross-Query Consistency Hypothesis, which posits that correct
answers grounded in genuinely relevant evidence maintain stable
confidence across semantically equivalent but syntactically diverse
queries, whereas noise-induced hallucinations exhibit unstable
confidence because their support often depends on spurious,
query-conditioned evidence patterns. Based on this hypothesis, we
develop CQC-RAG, a training-free framework that co-designs
controlled query-level diversity injection with cross-query
consistency evaluation, operating entirely on a shared document pool
without increasing the retrieval budget.

CQC-RAG rewrites the original question into meaning-preserving query
variants, constructs query-conditioned reasoning contexts through
reranking, and selects answers according to confidence stability
across these contexts. This design enables answer self-evaluation
without external supervision and does not rely on expanded retrieval
coverage. Experiments on open-domain QA benchmarks demonstrate that
CQC-RAG improves answer reliability over representative multi-path
RAG baselines, especially in complex multi-hop scenarios where
retrieval noise and biased evidence can be amplified. Our work
suggests a shift from finding a single optimal query toward
evaluating answer stability across semantically equivalent query
perspectives.

\appendix

\section{Prompt for Parallel Rewriting}

\subsection{Prompt for Parallel Rewriting on TrivialQA}

\begin{ragprompt}
Original query: \\
In Buddhism, what is the state of blissful repose or absolute
existence by someone relieved of the necessity of rebirth?
\promptsep{Prompt} You are an AI assistant designed to improve the
retrieval recall of a RAG system for a trivia question answering
task (TriviaQA).
Your goal is to generate 4 paraphrased versions of a given user query.\\

\#\#\# CRITICAL RULES:\\
1. **Preserve Entities:** NEVER change or omit proper nouns, dates, song titles, character names, or quoted text (e.g., "The Miami Sound Machine", "Dec 14, 1636", "Winter Wonderland"). These are the keys to finding the answer.\\
2. **Maintain Semantics:** The paraphrased questions must ask for EXACTLY the same answer as the original. Do not generalize or make the question vague.\\
3. **Vary Structure:**
   - Switch between Question formats (Who/What/Which) and Imperative formats (Identify/Name/State).
   - Switch between Active and Passive voice.
   - Move the focus of the sentence (e.g., put the date at the beginning or end).\\
4. **No Hallucination:** Do not add external information that is not present in the original question.\\

\#\#\# Few-Shot Examples:\\

**Example 1:**
Original: "Whose backing band is known as The Miami Sound Machine?"
Paraphrases: 1. "The Miami Sound Machine is the backing band for
which famous artist?" (Structure change) 2. "Identify the singer
cooperated with the band The Miami Sound Machine." (Imperative) 3.
"Which musician uses The Miami Sound Machine as their backing
group?" (Synonym: group vs band)
4. "Name the performer whose band goes by the name The Miami Sound Machine." (Focus shift)\\

**Example 2:**
Original: "According to the proverb, what comes but once a year?"
Paraphrases: 1. "What is it that arrives only one time annually,
based on the proverb?" (Synonym: once a year -> annually) 2.
"Identify the subject of the proverb stating it comes but once a
year." (Structure change) 3. "Which event is described in the
proverb as occurring only once a year?" (Question word change)
4. "According to the saying, what happens just a single time per year?" (Synonym: saying vs proverb)\\

**Example 3:**
Original: "With a motto of Always Ready, Always There, what US
military branch had it's founding on Dec 14, 1636?" Paraphrases: 1.
"Which branch of the US military was founded on Dec 14, 1636, and
uses the motto 'Always Ready, Always There'?" (Reordering) 2. "Name
the US military force established on Dec 14, 1636, with the slogan
'Always Ready, Always There'." (Synonym: branch -> force) 3. "The
motto 'Always Ready, Always There' belongs to which US service
branch founded in 1636?" (Focus on Motto)
4. "Identify the military entity formed on Dec 14, 1636, that says 'Always Ready, Always There'." (Imperative)\\

\#\#\# FORMAT\\
Output only the 4 paraphrased questions, one per line. No numbering,
no explanations. \promptsep{Output}
Paraphrased queries:\\
1. What is the state of blissful repose or absolute existence in Buddhism for someone who is no longer bound by the cycle of rebirth?\\
2. Identify the Buddhist state of blissful repose or absolute existence that is attained by someone freed from the necessity of rebirth.\\
3. Which Buddhist concept refers to a state of blissful repose or absolute existence for someone who has been relieved of the need for rebirth?\\
4. In Buddhism, what is the term for the state of blissful repose or
absolute existence that comes from being free of rebirth?
\end{ragprompt}

\subsection{Prompt for Parallel Rewriting on MuSiQue}

\begin{ragprompt}
Original query: \\
What month did the Tripartite discussions begin between Britain,
France, and the country where, despite being headquartered in the
nation called the nobilities commonwealth, the top-ranking Warsaw
Pact operatives originated? \promptsep{Prompt} You are an expert AI
assistant specializing in query rewriting for complex, multi-hop
reasoning tasks.
Your goal is to generate 4 paraphrased versions of a given user query to improve retrieval recall in a RAG system.\\

\#\#\# DATASET CONTEXT:\\
The queries come from the MuSiQue dataset, which involves multi-hop
reasoning. Questions often contain nested clauses, intermediate
entities, or multiple constraints that must be satisfied to find the
answer.

\#\#\# CRITICAL RULES:\\
1. **Preserve Entities \& Proper Nouns:** NEVER change, omit, or translate proper nouns (e.g., "La fida ninfa", "A Don", "Marlene McFly"). These are the anchors for retrieval.\\
2. **Maintain Strict Logic:** The paraphrased question must ask for the EXACT same logical target. Do not drop any constraint (e.g., if it says "in the winter of 2014", do not change it to "in 2014").\\
3. **Unpack Nested Structures:**
   - If the question involves a chain of relations, try to rephrase by changing the word order to highlight different keywords.
   - You may split long relative clauses into clearer modifier phrases.\\
4. **Variety in Expression:**
   - Use imperative forms (Identify/Name/Find).
   - Use interrogative forms (Who/What/Which).
   - Invert the sentence structure.

\#\#\# Few-Shot Examples:\\
**Example 1:**
Original: Who is the director of the film that features the song "My
Heart Will Go On"? Paraphrases: 1.Name the director of the movie in
which the song "My Heart Will Go On" appears. 2.The song "My Heart
Will Go On" was featured in a film directed by whom? 3.Identify the
filmmaker responsible for the movie containing "My Heart Will Go
On".
4.Who directed the motion picture that includes the track "My Heart Will Go On"?\\
**Example 2:**
Original: What is the capital of the country that shares a border
with the birthplace of Mozart? Paraphrases: 1.Identify the capital
city of the nation bordering Mozart's birthplace. 2.Mozart was born
in a city; which country's capital borders that location? 3.Name the
capital of the country adjacent to the place where Mozart was born.
4.Which city serves as the capital for the country bordering Mozart's birth city?\\
**Example 3:**
Original: The creator of the comic strip involving the character
Snoopy served in which branch of the military? Paraphrases: 1.In
which military branch did the creator of the Snoopy comic strip
serve? 2.Identify the branch of service of the man who created the
character Snoopy. 3.What military service did the author of the
comic featuring Snoopy perform?
4.Name the military branch associated with the creator of Snoopy's comic strip.\\

\#\#\# FORMAT\\
Output only the 4 paraphrased questions, one per line. No numbering,
no explanations. \promptsep{Output}
Paraphrased queries:\\
1. What month did the Tripartite discussions start between Britain, France, and the country where the top-ranking Warsaw Pact operatives came from, even though the headquarters of the nobilities commonwealth is located there?\\
2. In which month did the Tripartite discussions take place between Britain, France, and the nation that is the origin of the top-ranking Warsaw Pact operatives, despite its headquarters being in the nobilities commonwealth?\\
3. Figure out the month when the Tripartite discussions began between Britain, France, and the country where the top-ranking Warsaw Pact operatives originated, even though the headquarters of the nobilities commonwealth is situated there.\\
4. Which month marked the beginning of the Tripartite discussions
between Britain, France, and the country that is the birthplace of
the top-ranking Warsaw Pact operatives, despite the headquarters of
the nobilities commonwealth being located in that nation?
\end{ragprompt}

\section{Prompt for Multi-view Reasoning}

\begin{ragprompt}
\#\#\# Task Requirements:\\
            1. **Search First:** distinctively identify the sentence(s) in the context that directly support the answer.\\
            2. **Quote Verbatim:** You must extract the supporting text EXACTLY as it appears in the context.\\
            3. **Answer Second:** Formulate your answer based ONLY on the extracted evidence.\\
\#\#\# Output Format:\\
            Answer: [Your concise answer here]\\
            Evidence: [Paste the exact sentence from the context here]
\end{ragprompt}

\section{Artifacts}

All implementation artifacts used to produce the results reported in
this paper, including the CQC-RAG source code, preprocessed
datasets, and evaluation scripts, are publicly available at
\url{https://github.com/FrancesPlus/CQC-RAG}.

\bibliographystyle{ACM-Reference-Format}
\bibliography{sample-base}

\end{document}